\documentclass[showpacs,twocolumn,aps]{revtex4}
\usepackage{graphicx}
\usepackage{amsmath}
\usepackage{amssymb}
\usepackage{psfrag}

\newcommand{\dd}{\text{d}}

\newcommand{\ee}{\text{e}}

\newcommand{\p}{\partial}

\newcommand{\LAMB}{s}
\usepackage[latin1]{inputenc}
\usepackage[T1]{fontenc}
\usepackage{ae,aecompl}

\begin{document}
\title{Chaotic properties of systems with Markov dynamics} 
\author{V. Lecomte$^{1}$, C. Appert-Rolland$^2$  and F. van Wijland$^{1,3}$}
\affiliation{$^{1}$Laboratoire de Physique Th\'eorique 
(CNRS UMR8627), B\^atiment 210, 
Universit\'e Paris-Sud, 91405 Orsay cedex, France}
\affiliation{$^{2}$Laboratoire de Physique  Statistique 
(CNRS UMR8550), \'Ecole Normale Sup\'erieure, 
24 rue Lhomond, 75005 Paris, France}
\affiliation{$^{3}$Laboratoire Mati\`ere et Syst\`emes Complexes 
(CNRS UMR7057), Universit\'e Denis Diderot (Paris VII), 
2 place Jussieu, 75251 Paris cedex 05, France}
\begin{abstract}
  We present a general approach for computing the dynamic partition function
  of a continuous-time Markov process. The Ruelle topological pressure is
  identified with the large deviation function of a physical observable. We
  construct for the first time a corresponding finite Kolmogorov-Sinai entropy
  for these processes. Then, as an example, the latter is computed for a
  symmetric exclusion process. We further present the first exact calculation
  of the topological pressure for an $N$-body stochastic interacting system,
  namely an infinite-range Ising model endowed with spin-flip dynamics.
  Expressions for the Kolmogorov-Sinai and the topological entropies follow.
\end{abstract}
\maketitle

In statistical mechanics, bridging the microscopics to the macroscopics
remains the ultimate goal, be it in or out of equilibrium. The development of
the theory of dynamical systems and of their chaotic properties has led to
major advances in equilibrium and nonequilibrium statistical mechanics. All
those approaches make extensive use of such concepts as Lyapunov exponents,
Kolmogorov--Sinai (KS) or topological entropies, topological pressure, {\it
  etc.}, all quite mathematical in nature, and for which very few results
(even nonrigorous) are available, as far as systems with many degrees of
freedom are concerned.  One of the central ideas in constructing a statistical
physics out of equilibrium is that of Gibbs ensembles~\cite{ruelle} in which
time is seen to play the r\^ole of the volume in traditional equilibrium
statistical mechanics.  A central quantity called the dynamical partition
function is in general defined as
\begin{equation}
Z(\LAMB,t)=\!\!\!\!\!\!\!\!\!\!\!\sum_{\text{histories from }0\to t}\!\!\!\!\!\!\!\!\!\!\!\big(\text{Prob}\{\text{history}\}\big)^{1-\LAMB}
\end{equation}
The so-called thermodynamic formalism allows to derive from this quantity a
number of chaotic properties such as the topological entropy which is defined
through the number of possible trajectories a system can follow, or the KS
entropy which is a measure of the complexity of a process and characterizes
its dynamical randomness.  Within that framework there exists a mathematical
construction of smooth stationary measures for nonequilibrium steady-states
--the Sinai-Ruelle-Bowen (SRB) measures-- the determination of which precisely
rests on the dynamic partition function and the KS entropy~\cite{beckschlogl}.

It must however be acknowledged that the determination of 
any of these quantities has only been carried out for toy models
of dynamical systems, such as the baker's map~\cite{dorfman}. Most of the
efforts for physically relevant systems have borne on the Lorentz
gas~\cite{appertvanbeijerenernstdorfman, vanbeijerendorfman,
  dorfmanernstjacobs}, and more recently on hard-sphere systems in the dilute
limit~\cite{vanbeijerendorfmanposchdellago}. It is possible to relate
discrete-time Markov processes to dynamical systems (see
\cite{dorfmanernstjacobs} for a pedagogical account), yet very few examples
going beyond the simple random
walk have been investigated. More importantly, when the discrete time scale is
sent to zero, infinities arise~\cite{gaspard0}, so that no viable definition
of the dynamical partition function in the physical limit of continuous time
has hitherto been proposed. Continuous-time Markov processes are ubiquitous.
Many systems governed by a Hamiltonian dynamics can be mapped, within some
well-controlled approximation schemes, onto Markov processes. An extensive
activity in the physical modeling of complex systems, from interacting spins
to interacting gases, from avalanches in sandpiles to chemical reactions,
relies on a continuous-time Markov description.  In the present letter we
construct, apparently for the first time in the literature, the dynamical
partition function for such Markov processes. We do not rely on a
discrete-to-continuum limit~\cite{gaspard0}, and by working directly with
continuous time, we obtain finite results. From this we shall show how to
extract the KS entropy. We will further connect it, in the spirit of
Gaspard~\cite{gaspard1}, to the Markov analog of a fluctuating entropy flow,
as identified by Lebowitz and Spohn\cite{lebowitzspohn}. In order to exemplify
our findings, we will then perform an exact calculation on the infinite-range
({\it i.e.} mean-field) Ising model, endowed with spin-flip dynamics as
introduced by Ruijgrok and Tjon~\cite{ruijgroktjon}. To the best of our
knowledge, this is the first available result for a many-body interacting
stochastic system.

The outline of this letter is as follows: First we construct the dynamical
partition function and the related topological pressure, which is identified
as the large deviation function of a physical observable. From this we deduce
an expression for the KS entropy. Then we establish a connection to the
time-integrated entropy current. Finally, we explore the consequences of our
formulation on three examples, a random walk with absorbing boundaries, a
symmetric exclusion process, and an infinite-range
Ising model, for which explicit and exact calculations are performed.\\

We begin with a generic Markov process characterized by transition rates
$W({\cal C}\to{\cal C}')$ from configuration $\cal C$ to configuration ${\cal
  C}'$. The mean residence time in configuration $\cal C$ is $1/r({\cal C})$,
with $r({\cal C})=\sum_{{\cal C}'}W({\cal C}\to{\cal C}')$. This means in
particular that the probability of hopping from configuration $\cal C$ after a
time interval $t$ to some other configuration between $t$ and $t+\dd t$ is
$r({\cal C})\exp(-r({\cal C})t)\dd t$. Among the allowed target
configurations, the system hops to the particular configuration ${\cal C}'$
with probability $W({\cal C}\to{\cal C}')/r({\cal C})$. Various properties of
the master equation evolution operator $\mathbb{W}$ for the probability
$P({\cal C},t)$ to be in state ${\cal C}$ at time $t$, such that, in matrix
notation, $\p_t P=\mathbb{W}P$, can be found in \cite{vankampen}. We define
the dynamic partition function $Z(\LAMB,t)$ as the sum over all possible
histories of the process over the interval $[0,t]$ of the probabilities of the
histories raised to the power $(1-\LAMB)$~\cite{notationconflict}. It is a
matter of carefully applying the definition to realize that
\begin{equation}\label{dynZ}
Z(\LAMB,t)=\!\!\!\!\!\!\!\!\!\!\!\sum_{\text{histories from }0\to t}\!\!\!\!\!\!\!\!\!\!\!\big(\text{Prob}\{\text{history}\}\big)^{1-\LAMB}=\langle\ee^{-\LAMB Q_+(t)}\rangle
\end{equation}
where the observable $Q_+(t)$ depends on the sequence of states  ${\cal
C}_0,...,{\cal C}_k$ occupied by the system over $[0,t]$ through the
relationship
\begin{equation}
Q_+(t)=\sum_{n=0}^{k-1}\ln \frac{W({\cal C}_n\to{\cal C}_{n+1})}{r({\cal C}_n)}
\end{equation}
It is necessary to describe the meaning of the brackets in (\ref{dynZ}): they
stand for an average over the number $k$ of successive states occupied over $[0,t]$, over
the various configurations ${\cal C}_0,...{\cal C}_k$ visited by the system and,
finally, over the time lapses that the system has been staying in each of
those $k$ states. All these quantities define a {\it history}.
Note that the topological or Ruelle pressure $\psi(\LAMB)$, canonically defined as
$\psi(\LAMB)=\lim_{t\to\infty}\frac{1}{t}\ln Z$, is also the generating
function of the cumulants of the physical observable $Q_+$. It is possible to
write an evolution equation for $P({\cal C},Q_+,t)$, the probability to be
in state $\cal C$ at time $t$ with $Q_+(t)=Q_+$, and a similar one for the
related generating function $\int\dd Q_+\ee^{-\LAMB Q_+}P({\cal C},Q_+,t)$.
The latter obeys a master-equation-like evolution with an operator
$\mathbb{W}_+$ whose matrix elements are given by
\begin{equation}\label{defW+}
\mathbb{W}_+({\cal C},{\cal C}')=W({\cal C}'\to{\cal C})^{1-\LAMB}
r({\cal C}')^\LAMB-r({\cal C})\delta_{{\cal C},{\cal C}'}
\end{equation}
An important
consequence is that $\psi(\LAMB)$ is simply the largest eigenvalue of the
evolution operator $\mathbb{W}_+$.
The analog was already known~\cite{gaspard0} for a
discrete-time Markov process, but appears to be a new result for
continuous-time processes. We stress that $\psi$ is finite even though we are
working in continuous time. Having reduced the computation of chaotic
properties to a mere technical challenge, we shall later present concrete
physical examples in which the entire spectrum of $\mathbb{W}_+$ can be
determined. We temporarily continue with abstract considerations by defining a
similar quantity $Q_-$ for the time-reversed process,
\begin{equation}
Q_-(t)=\sum_{n=1}^{k}\ln\frac{W({\cal C}_{n}\to{\cal C}_{n-1})}{r({\cal C}_{n-1})}
\end{equation}
which verifies that
\begin{equation}\label{lienQ+Q-QS}
Q_+(t)-Q_-(t)=\sum_{n=1}^{k}\ln\frac{W({\cal C}_{n-1}\to{\cal C}_{n})}{W({\cal
C}_{n}\to{\cal C}_{n-1})}=Q_S(t)
\end{equation}
The quantity $Q_S(t)$ appearing in the rhs of (\ref{lienQ+Q-QS}) is precisely
the time-integrated fluctuating entropy flow introduced by Lebowitz and
Spohn~\cite{lebowitzspohn}, which verifies, as $t\to\infty$ the
Gallavotti-Cohen symmetry $\langle\ee^{-s Q_S}\rangle\simeq \langle
\ee^{-(1-s)Q_S}\rangle$. A similar property for discrete time Markov
processes, discussed at the level of the averages, was already noted by
Gaspard~\cite{gaspard1}. In the same way as it can be seen that the entropy
current
\begin{equation}
J_S({\cal C})=\sum_{{\cal C}'}W({\cal C}\to{\cal C}')
\ln\frac{W({\cal C}\to{\cal C}')}{W({\cal C}'\to{\cal C})}
\end{equation}
verifies $\dd \langle {Q_S}\rangle/\dd t=\langle J_S({\cal C})\rangle$, one
can verify that
\begin{equation}J_\pm({\cal C})=\sum_{{\cal C}'}W({\cal C}\to{\cal C}')\ln\frac{W({\cal
C}\stackrel{\longrightarrow}{\leftarrow}{\cal C}')}{r({\cal C})}
\end{equation}
govern the evolutions of $Q_\pm$ according to $\dd \langle {Q_\pm}\rangle/\dd
t=\langle J_\pm({\cal C})\rangle$, and they have the property that $J_+({\cal
  C})-J_-({\cal C})= J_S({\cal C})$. The entropy current is usually
proportional to a particle current~\cite{bodineauderrida,derridalebowitz,
  bertinidesolegabriellijonalasiniolandim} or to an energy current
~\cite{lecomteraczvanwijland}, hence its physical meaning is clear. However
neither of $J_\pm$ receives such a straightforward interpretation. In
equilibrium, that is when the rates $W$ satisfy the detailed balance
condition, both $Q_+$ and $Q_-$ have the same large deviation function. It
requires to relate $Q_+$ to quantities familiar in the dynamical systems
theory to endow it with a more transparent physical meaning.  The KS entropy,
$h_{\text{KS} }$, which is defined~\cite{beckschlogl} as
\begin{equation}\label{defhks}
h_{\text{KS} }=-\frac{1}{t}\sum_{\text{hist.}}\text{Prob}\{\text{history}\}\ln
\text{Prob}\{\text{history}\}
\end{equation}
is also obtained,
for a system without escape, from
\begin{equation}\label{}
h_{\text{KS} }=\psi'(0)=-\langle J_+\rangle.
\end{equation}
Therefore $J_+$ appears to convey the physical meaning of an information
content flow. Another quantity of interest is the topological entropy
$h_{\text{top}}=\psi(1)$, which counts the number of possible trajectories
over $[0,t]$. We close our construction with a remark on SRB measures.
Following the procedure outlined in Beck and Schl\"ogl~\cite{beckschlogl}, the
stationary SRB measure can be obtained from a variational principle: in our
case, this is the invariant measure $P$ that renders the combination
$\psi_{P}(\LAMB)=h_{\text{KS}}[P]-(1-\LAMB)\langle Q_+\rangle_{P}/t$ maximum
when $\LAMB=0$. The master equation possesses a unique stationary solution.
The SRB measure can only be this stationary solution, for which the above
combination becomes $\psi(\LAMB)=\LAMB \;h_{\text{KS}}$ if $\LAMB\to 0$ and
one recovers
the fact that $h_{\text{KS}}=\psi'(0)$.\\

We now present three simple applications of increasing complexity.  Consider
first a particle diffusing with diffusion constant $D$ on a one-dimensional
infinite line.  We can illustrate on this example how a deterministic map
allows to define for a stochastic system a Lyapunov exponent compatible with
the thermodynamic formalism exposed above.  Each jump of the particle can be
described by a deterministic map as indicated in
\cite{dorfmanernstjacobs,dorfman}.
This allows to define a corresponding Lyapunov exponent from the exponential
divergence between initially close trajectories, averaging over the time
lapses between jumps (which are randomly distributed as for the Markov
process) .  Here we find $\lambda = 2 D \ln 2$ which is related to the
topological pressure $\psi(\LAMB) = 2D\left(2^\LAMB -1\right)$ through
$\psi^\prime(0) = \lambda$.

The same example can be used to illustrate the case of systems with escape.
The particle now jumps on an infinite two-dimensional lattice slab of width
$\ell$ with absorbing boundaries. It is a trivial matter to diagonalize the
corresponding $\mathbb{W}_+$ and to find its largest eigenvalue $\psi(\LAMB)$,
which reads, for large $\ell$
\begin{equation}
\psi(\LAMB)=4D(4^\LAMB-1)-4^\LAMB D\pi^2/\ell^2
\end{equation}
Hence the topological pressure verifies $\psi(0)=-\gamma$, where
$\gamma=D\pi^2/\ell^2$ is the escape-rate of the particle. This result is of
course consistent with that established for the Lorentz
gas and stands as yet another manifestation of the link
between transport coefficients and the topological
pressure~\cite{gasparddorfman}.

Our second example is the {\it Symmetric Exclusion Process}, a gas of $N$
mutually excluding particles diffusing on a one-dimensional lattice of $L$
sites with periodic boundary conditions. Their hopping rate is denoted
by $D$, and for simplicity we have confined our analysis to determining the
KS entropy. Denoting by $\rho$ the average density, and by
$\sigma(\rho)=2\rho(1-\rho)$ (twice) its compressibility, we find that
\begin{equation}\label{hksSEP}
h_{\text{KS} }/D=L\sigma \ln (L\sigma)+
\sigma \ln (L\sigma)+\frac 32 \sigma+{\cal O}(\ln L/L)
\end{equation}
In order to establish this result we started from $h_{\text{KS} }=-\langle
J_+\rangle$ and we exploited that the equilibrium state is perfectly random
\cite{schutz}.  Our first comment on (\ref{hksSEP}) is that $h_{\text{KS} }$
is not extensive in the system size. Second, the effect of the interaction is
felt through a given combination of the system size and of its
compressibility. At half-filling the system shows its largest KS entropy, and
this is likely related to the fact that also the number of allowed distinct
microscopic states is maximum at $\rho=1/2$. The dependence of
$h_{\text{KS} }$ on $\rho$ through $\sigma$ only arises from the particle-hole
symmetry.

As a third example, we consider an infinite range Ising model with Hamiltonian
${\cal H}=-(\sum_{i=1}^N\sigma_i)^2/2N$, endowed with spin-flip dynamics. Each
spin $\sigma_i$ flips independently with a rate $\exp(-\beta \sigma_i M/N)$,
where $\beta$ is the inverse temperature and where $M=\sum_i\sigma_i$ is the
total magnetization before the spin-flip. These flipping rates satisfy the
detailed balance property with respect to the canonical distribution
$\exp(-\beta {\cal H})$. This system has the advantage of exhibiting a
second-order phase transition, the generic effect of which on Lyapunov
exponents has hitherto never been explicitly probed. Technically speaking, the
master equation has an evolution operator that can be exactly
diagonalized~\cite{ruijgroktjon}. We will now describe, skipping all technical
details~\cite{nous}, our results for the topological pressure for this system.
On the one hand we may follow the procedure outlined in the first part of this
letter, and identify a state $\cal C$ with a configuration $\{\sigma_i\}$ of
the $N$ spins. There are $2^N$ such configurations, and restricting ourselves
to the high temperature phase $\beta< 1$, the exact diagonalization of
$\mathbb{W}_+$ 
as defined in (\ref{defW+}) 
leads to the eigenvalues (or
Ruelle-Pollicott resonances) $\psi(\LAMB)-n\phi(\LAMB)$ with $n\in\mathbb{N}$,
where
\begin{eqnarray}\nonumber\label{psiIsing}
\psi(\LAMB)=(N^\LAMB-1)N+N^\LAMB(1-(1-\LAMB)\beta)-\phi(\LAMB)\\
\phi(\LAMB)= N^{\LAMB/2}\sqrt{N^\LAMB(1+\LAMB\beta(2-\beta))-\beta(2-\beta)}
\end{eqnarray}
Again we remark that the topological pressure
(\ref{psiIsing}) is not extensive in the number of spins, which also reflects
on the KS entropy:
\begin{equation}\label{hksIsing}
h_{\text{KS}}=N\ln N-\frac{\beta(2-\beta)}{2(1-\beta)}
\ln N-\frac{\beta^2}{2(1-\beta)}
\end{equation}
What can also be observed on (\ref{hksIsing}) is that, due to the diverging
susceptibility $\beta/(1-\beta)$ at the critical point $\beta=1$ our
calculational technique~\cite{ruijgroktjon} ceases to be valid for systems
with $N(1-\beta)\sim 1$.

Another viewpoint, on the other hand, would have consisted in adopting a
coarse-grained description of the same spin system and chosing to characterize
its states by their total magnetization $M$ (there are $2N$ such magnetization
states). Even though we are talking about the same system we are not talking
about the same Markov process, hence the spectrum of $\mathbb{W}_+$ differs
from its previous expression and the Ruelle pressure is now given by
\begin{equation}\begin{split}
{\psi}(\LAMB)=&N(2^\LAMB-1)+2^\LAMB(1-\LAMB)(1-\beta)\\
&-2^{\LAMB/2}
\left[2^\LAMB(1-\LAMB(1-\beta)^2)-\beta(2-\beta)\right]^{1/2}
\end{split}\end{equation}
Now the corresponding KS entropy is extensive in the system size. The leading
term in the system size expresses that, roughly speaking, the total
magnetization undergoes a simple random walk. Interactions, and the effects of
correlations, just as in (\ref{hksSEP}) or (\ref{hksIsing}), are felt to the
next order in the system size. This is where interesting physical features are
to be looked for.

We now summarize our findings. We have defined and constructed a dynamical
partition function {\it à la} Ruelle, along with the related topological
pressure for continuous-time Markov processes. The topological pressure
appears to be the largest eigenvalue of some operator, and is therefore a
finite quantity, in spite of working within a continuous time formulation. A
key quantity for characterizing dynamical randomness, the KS entropy, follows,
along with the topological entropy and the sum of positive Lyapunov exponents.
We have put our definitions to work on a few simple examples, thus
illustrating how straightforward the connection between topological pressure
and escape rate is, and making a significant step towards realistic systems
(with large number of degrees of freedom and interactions). Our formulation
has the advantage of bringing concepts and quantities pertaining to other
fields of studies --dynamical systems theory-- back into the statistical
mechanics community, and to which the available toolbox (simulations,
mean-field approximations, low density expansions, field theory, exact
results, {\it etc.}) will apply. We insist that the observable $Q_+(t)$ from
which the main dynamical quantities arise can be easily monitored when
performing numerical simulations.

We believe that our formulation opens a series of new research routes. To
begin with, it would be interesting to see, {\it e.g.} for a lattice gas, how
the precise form of the interactions and of the microscopic dynamics affects
the KS entropy, and possibly to help clarify ongoing
debates~\cite{lienhks-diff}. To what extent is the result obtained in
(\ref{hksSEP}) universal? Next, it would be instructive to see if
ergodicity-breaking features, which are suspected to characterize systems with
glassy dynamics, can be identified on $h_{\text{KS}}$. This could be first
examined on simple systems like a particle diffusing in a random scenery, and
then on a many-body glass forming lattice gas, such as the Kob-Andersen model.
Even if no conceptual progress is to be expected from that, it would certainly
be useful to specialize our approach to a widely used class of Markov
processes, that which rests on Langevin equations (this could be achieved by
resorting to the description used by Kurchan in \cite{kurchan}). This includes
in particular fluctuating hydrodynamics, but also phenomenological models of
chemical reactions, surface growth, turbulence, {\it etc}. It would also be
interesting to see how the nonequilibrium nature can be traced back on the
form and properties of the topological pressure. 
In that sense it would be useful to compare the dynamical entropies of a gas
in equilibrium and for the same gas driven out of equilibrium by the system
boundaries or by a bulk field.  Such a system develops long range correlations
and, in the case of a bulk drive, nonequilibrium phase transitions may be
observed~\cite{schutzdomany,derridaevanshakimpasquier}; how these features
affect the dynamical entropies deserves to be investigated.
 Finally, the link with dynamical system
theory that we have presented in our first example still involves a stochastic
part (the residence time in each visited state).  It would be interesting to
formalize a relationship between continuous-time Markov processes and fully
deterministic flows.  This identification, if possible, would perhaps
allow for an interpretation of our KS entropy in terms of the sum of Lyapunov
exponents for the corresponding flow. We hope that the present work will stand
as a contribution towards welding communities somewhat impervious to each
other.\\
\noindent {\it Acknowledegments.--} The authors thank M. Courbage, H.J. Hilhorst,
J.R. Dorfman and H. van Beijeren for useful
discussions and for critical comments.


\begin{thebibliography}{99}
\bibitem{notationconflict}
We have substituted $1-\LAMB$ to the mathematical notation $\beta$, which we
reserve for denoting an inverse temperature.  Besides, this notation is more
natural when $Z(\LAMB,t)$ is rewritten as a generating function.
\end{thebibliography}
\end{document}